\documentclass[aps,prd,floats,showpacs,nofootinbib,superscriptaddress,preprint,tightenlines]{revtex4}

\usepackage{graphicx}
\usepackage[centertags]{amsmath}

\def\bea{\begin{eqnarray}}
\def\eea{\end{eqnarray}}
\def\be{\nopagebreak[3]\begin{equation}}
\def\ee{\end{equation}}
\def\ba{\nopagebreak[3]\begin{eqnarray}}
\def\ea{\end{eqnarray}}

\def\N{{\cal N}}
\def\l{\lambda}
\usepackage{enumerate}

\usepackage{colordvi}

\begin{document}
\preprint{\vbox{\baselineskip=12pt \rightline{IGC-11/2-2}}}

\title{Measure problem in slow roll inflation and loop quantum cosmology}
\author{Alejandro Corichi}\email{corichi@matmor.unam.mx}
\affiliation{Instituto de Matem\'aticas,
Unidad Morelia, Universidad Nacional Aut\'onoma de
M\'exico, UNAM-Campus Morelia, A. Postal 61-3, Morelia, Michoac\'an 58090,
Mexico}
\affiliation{Center for Fundamental Theory, Institute for Gravitation and the Cosmos,
Pennsylvania State University, University Park
PA 16802, USA}
\author{Asieh Karami}
\email{karami@matmor.unam.mx} 
\affiliation{Instituto de F\'{\i}sica y
Matem\'aticas,  Universidad Michoacana de San Nicol\'as de
Hidalgo, Morelia, Michoac\'an, Mexico}
\affiliation{Instituto de Matem\'aticas,
Unidad Morelia, Universidad Nacional Aut\'onoma de M\'exico,
UNAM-Campus Morelia, A. Postal 61-3, Morelia, Michoac\'an 58090,
Mexico}

\begin{abstract}
We consider the measure problem in standard slow-roll inflationary models from the perspective
of loop quantum cosmology (LQC). 
Following recent results by Ashtekar and Sloan, we study the probability of having enough
e-foldings and focus on its dependence on the quantum gravity scale, including the transition 
of the theory to the limit where general relativity (GR) is recovered. 
Contrary to the standard expectation, the probability of having enough inflation, that is close to one in LQC,
grows and tends to 1 as one approaches the GR limit. We study the origin of the tension between these results with those by Gibbons and Turok, and offer an explanation that brings these apparent contradictory results into a coherent picture. As we show, the conflicting results stem from different choices of initial conditions for  the
computation of probability. 
The singularity free scenario of loop quantum cosmology offers a natural choice of initial conditions, and suggests that enough inflation is generic.
\end{abstract}

\maketitle

\section{Introduction}

The measure problem in cosmology has received some attention since it was suggested that one should weight, over the space of classical solutions to the equations of general relativity, those solutions that exhibit enough inflation to account for present observations \cite{haw:page}. An early observation was that there exists a natural measure on the phase space of the theory with respect to which one should compute probabilities. Recently, Gibbons and Turok overcame some early difficulties in the total normalization and concluded that, for the simplest inflationary potentials in a FRW universe, the probability of inflation was greatly suppressed \cite{gibb:turok}. One potential difficulty with such calculations pertains to the choice of initial conditions. Since all solutions to the equations of motion are singular in the past (for expanding universes), one needs a prescription for selecting initial conditions for those solutions. In  \cite{gibb:turok} such a prescription was put forward in terms of a `constant density surface', roughly speaking, at the end of inflation. Another possibility is given by defining a `cut-off', in the form of a constant density surface at, say, the Planck scale, as was early suggested in \cite{BKGZ}. 

Yet another possibility is that a quantum theory of gravity might be able to provide such Planck surface in a natural way. Such is indeed the case of loop quantum cosmology \cite{lqc}, a quantum framework closely related to loop quantum gravity \cite{lqg} that has been able to achieve robust results regarding avoidance of big bang singularities \cite{aps,slqc} (See, for instance, \cite{AA} for a recent survey). In LQC, all trajectories undergo a {\it bounce} that replaces the initial singularity, attain a maximum {\it critical density} \cite{slqc}, and preserve semiclassicality across the bounce \cite{recall}, thanks to uniqueness results that warranty the consistency of the theory \cite{uniqueness}.  
Two key results in the measure problem have been obtained in LQC. First, it has been shown that one could account for the dynamics of the quantum universe by means of {\it effective} equations that capture the main quantum gravity effects and that reduce to the classical equations in the appropriate regime \cite{victor,SVV}. This was used in \cite{SVV} to show that, for several inflationary potentials,  the characteristic `attractor behavior' of inflationary dynamics  \cite{BKGZ,kofman,linde} is recovered in the low energy regime.
Furthermore, Ashtekar and Sloan showed recently that the natural measure of \cite{gibb:turok} can be finitely implemented in LQC, and proposed a natural Planck scale surface on which to compute probabilities \cite{ash:sloan}. Surprisingly, the probability for having enough e-foldings was shown to be close to one, in contrast to the result of Gibbons and Turok that was done for classical GR\footnote{There have been several previous attempts to study
the issue of inflation within LQC. In \cite{germani} the natural measure of \cite{gibb:turok} was considered but
the effects of the bounce and superinflation were ignored. In \cite{miel} the issue of the measure was not considered and only a small part of the parameter space was explored.}.

In loop quantum cosmology, the underlying discreetness of the quantum geometry manifests itself via a dimension-full parameter $\lambda$. In the LQC literature it is standard to choose the value of $\lambda$ such that the minimum quantum of area corresponds to that found in LQG \cite{aps,lqg}. But, if one considers this as a free {\it phenomenological} parameter of the theory, it is natural to ask whether in the limit $\lambda \to 0$, where the loop quantum geometric effects disappear, 
one can recover the standard Wheeler-DeWitt quantum cosmology. This has been answered with different levels of sophistication \cite{aps,slqc,cvz3}. The authors of \cite{aps} showed that the difference equation governing the LQC
dynamics reduces to the differential WDW equation in the large volume limit. Later, in \cite{slqc} and \cite{cvz3}, the limit $\lambda \to 0$  was studied and it was shown that one does recover the standard WDW and the GR limit in some regime. In the case of {\it effective} classical equations, in this limit one recovers the equations of general relativity. 

The purpose of this article is to explore the relation between loop quantum cosmology and general relativity, as we take the limit $\lambda \to 0$, regarding the measure problem in slow roll inflation. More precisely, we would like to understand the apparent tension between the results of Gibbons and Turok, with those of Ashtekar and Sloan. If one starts with the analysis of \cite{ash:sloan}, that was done for a fixed value of $\lambda$ (of the order of the Planck scale), and one takes the limit $\lambda \to 0$, one might expect to recover the results of Gibbons and Turok. As we shall show in detail this expectation is not realized. Indeed, quite the opposite occurs. As the value of the discreetness parameter is decreased, the probability of having enough inflation {\it increases} and approaches one in the limit.
One would then be forced to conclude that in the general relativity limit of loop quantum cosmology, the probability of having enough inflation is (almost) one, in stark contrast with the analysis of Gibbons and Turok. 

What is then the source of this
apparent tension? As we shall argue, the tension is resolved once one analyzes in detail the assumptions underlying both calculations. The difference turns out to be due to the initial conditions one imposes on the corresponding
`constant density surface'. In the Gibbons and Turok analysis this is taken near the end of inflation, well below the Planck scale, whereas in the LQC calculation one is taking it at the scale set by the parameter $\lambda$ (which in the Ashtekar and Sloan analysis is close to the Planck scale). In the limit $\lambda \to 0$ the energy density
at which the initial conditions are defined in LQC diverges, so one comes closer to the big bang singularity 
as one approaches the GR limit. It is this difference what accounts for the conflicting conclusions.

The structure of the paper is as follows. In Sec.~\ref{sec:2} we give a brief review of the effective description for loop quantum cosmology of a $k$=0 FRW cosmology with a scalar field. In Sec.~\ref{sec:3} we present the calculation of the probability for having $\N$ e-foldings or more in LQC. We put special attention to the discreetness parameter of LQC and the limit when it vanishes. Next, we give an argument based on global properties of the dynamics and the Liouville measure to understand the results of both \cite{ash:sloan} and \cite{gibb:turok}. We end in Sec.~\ref{sec:4} with a discussion. Throughout the paper we use Planck units, where $G$=$\hbar$=$c$=1, (rather than $8\pi G=$1, a  convention sometimes used in cosmology).

\section{Effective dynamics in Loop Quantum Cosmology}
\label{sec:2}

Let us now give a brief review of the effective formalism in LQC.
The effective Hamiltonian that one obtained from loop quantum cosmology for  a  $k$=0  FRW model is \cite{victor}
\be
\mathcal{H}_{\textrm{eff}}=-\frac{3}{8\pi \gamma^2\lambda^2}v\sin^2{\lambda\beta}+\rho v
\label {h}
\ee
where $v$ is the volume and, on equations of motion, $\beta=\gamma H$, where $H$ is the Hubble parameter. From the previous Hamiltonian the {\it effective} Friedman equation becomes,
\be
\frac{\sin^2{\lambda\beta}}{\gamma^2\lambda^2}=\frac{8\pi}{3}\rho
\label{f1} 
\ee
or, equivalently
\be
H^2=\frac{8\pi}{3}\rho\left(1-\frac{\rho}{\rho_{\textrm{crit}}}\right)
\label{f2}
\ee
where the density is given by $\rho=\dot\phi^2/2+V(\phi)$. Here $\rho_{\textrm{crit}}=3/(8\pi\gamma^2\lambda^2)$, the {\it critical} density, is the density of the scalar field at the bounce. All trajectories undergo a bounce for which the density becomes exactly $\rho_{\textrm{crit}}$.
 In the low density regime, namely when 
$\lambda\beta\ll 1$ or $\rho\ll\rho_{\textrm{crit}}$ we approach classical general relativity. Note that the quantum geometry scale $\lambda$ sets the scale for the critical density. With the standard value taken in the LQC literature $\lambda=\sqrt{4\sqrt{3}\pi\gamma}\,\ell_{\textrm Pl}$ and the Barbero-Immirzi parameter $\gamma \approx 0.237$ chosen to be compatible with the Hawking-Bekenstein entropy \cite{bh}, the critical density is $\rho_{\textrm{crit}}\approx 0.41\rho_{\textrm Pl}$ \cite{slqc}. (Recall that is the Planck units we are using $\ell^2_{\textrm Pl}$=$G\hbar$=1, and $\rho_{\textrm Pl}=1$.) As we decrease the parameter $\lambda$, the critical density increases, so the `classical limit' is attained in the limit when the critical density diverges.

\begin{figure}[tb]
\centerline{\includegraphics [scale=0.4]{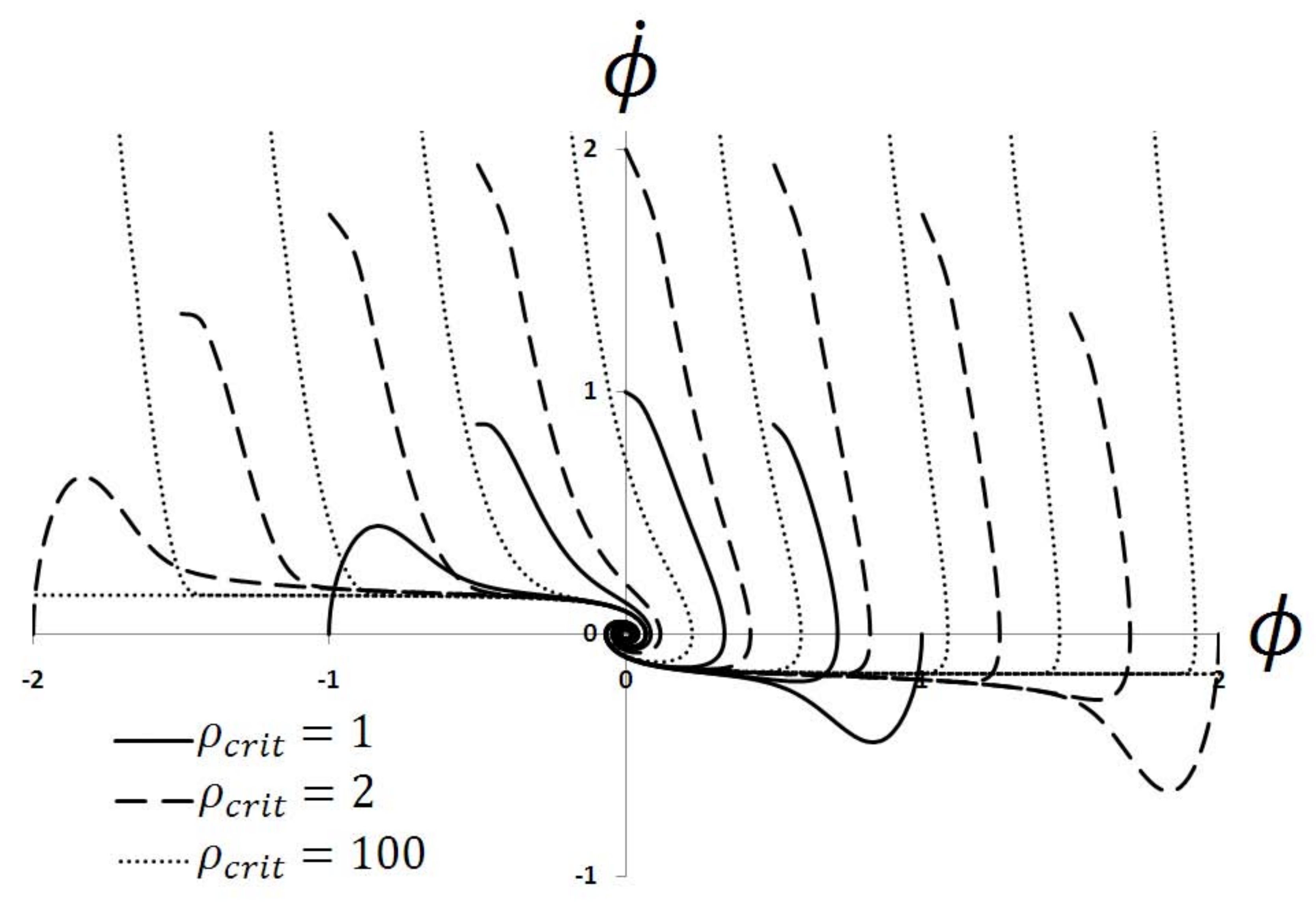}}
\caption{Three sets of trajectories are plotted for different values of the critical density. Note that near the origin, all trajectories approach the attractor.}
\label{Fig:1}
\end{figure}
The equation of motion for the scalar field $\phi$ yield the standard Klein Gordon equation is,
\be
\ddot\phi+3H\dot\phi+V_{,\phi}=0
\label{phi}
\ee
For the simplest potential, namely $V=m^2\phi^2/2$, we have solved the equations of motion for various values of the critical density and for convenience, plotted them in Fig.~\ref{Fig:1}. In the $(\phi,\dot{\phi})$ plane, the surfaces of constant density are ellipsoids defined by $\rho=\dot{\phi}^2/2 + m^2\phi^2/2$. All trajectories approach the `critical density surface', the ellipse bounding the phase diagram where the bounce occurs and touch it tangentially. Something that one might expect and that was checked in \cite{SVV}, is that near the origin of the plane, where the density is small compared to the critical one, the LQC trajectories and the classical one should coincide. This can be seen in Fig.~\ref{Fig:1}. As one decreases $\lambda$ the critical density increases and the maximum ellipse
defined by $\dot{\phi}^2_{B}/2 + m^2\phi_{B}^2/2=\rho_{\textrm{crit}}$ becomes larger.
The classical limit (GR) can be approached as $\lambda\to 0$. One has to note however, 
that this limit is somewhat discontinuous
\cite{slqc,cvz3}, since all LQC trajectories bounce, for all values of $\lambda$, while there is no
bounce in GR. In this particular sense, the GR `limit', and correspondingly the big bang, 
corresponds to an `infinitely large ellipsoid', or the point at infinity in the $(\phi,\dot\phi)$ plane (See Fig.~\ref{Fig:3}).

\section{Probability for Slow Roll Inflation in LQC}
\label{sec:3}

This section has two parts. In the first one, we calculate the probability for slow roll inflation in LQC and consider the limit when the discreetness parameter tends to zero. In the second part, we use qualitative aspects of the dynamics to gain a deeper understanding of the results.

\subsection{Probability}

\par
Let us now evaluate the probability for inflation as done in \cite{ash:sloan}, keeping track of the dependence on $\lambda$. Without loosing generality, for the remainder of the article we shall focus on the sector of the solution space for which $\dot\phi$ is nonnegative. Then, the Liouville measure $\textrm{d}\mu$ when pulled back to
the surface with constant $\beta$ or equivalently with constant $\rho$, has the from \cite{ash:sloan},
\be
\textrm{d}\mu=\sqrt{8}\pi\gamma\, v(\rho-V(\phi))\,\textrm{d}\phi\,\textrm{d}v
\label{lm}
\ee
We we further choose, as in \cite{ash:sloan}, the surface of constant $\beta$ (and $\rho$) at the bounce, we get
\be
\textrm{d}\mu=\frac{\sqrt{3\pi}}{\lambda}\, v_B\sqrt{1-F_B}\,\textrm{d}\phi_B\textrm{d}v_B
\label{lmb}
\ee
where $\phi_B$ is the value of scalar field at the bounce, $v_B$ is the volume of the universe at the bounce and $F_B=V(\phi_B)/\rho_{\textrm{crit}}$. This is the measure that will be used for computing the probability 
of having $\N$ or more e-foldings.

The number of e-foldings during inflation, $\mathcal{N}$, can be written as
\be
\mathcal{N}=\int_{t_o}^{t_{\textrm{end}}} H\,\textrm{d}t=\int_{\phi_o}^{\phi_{\textrm{end}}}\frac{H}{\dot\phi}\,\textrm{d}\phi
\ee
where $t_o$, $\phi_o$, $t_{\textrm{end}}$,  and $\phi_{\textrm{end}}$ are the time and value of the scalar field at the onset and at the end of inflation, respectively. 
We can use the slow roll conditions, $V(\phi)\gg \dot\phi/2$ and
$V_{,\phi}\gg\ddot\phi$, together with Eq.({\ref{phi}) to approximate $\N$,
\be
\mathcal N\approx-\int_{\phi_o}^{\phi_{\textrm{end}}}\frac{3H^2}{V_{,\phi}}\,\textrm{d}\phi=2\pi\left(1-\frac{\phi_o^2+\phi_{\textrm{end}}^2}{2\phi^2_{\textrm{max}}}\right)(\phi_o^2-\phi_{\textrm{end}}^2)\,
\ee
where $\phi_{\textrm{max}}$ is the maximum value the scalar field can
attain and is given by
$\phi_{\textrm{max}}=\sqrt{2\rho_{\textrm{crit}}}/m$. For large
values of $\mathcal N$, the value of the scalar field at the end of
inflation is much smaller than its value at the onset of
inflation. Thus, for large (but finite) $\mathcal N$  we can neglect
some terms and get,
\be
\mathcal N\approx
2\pi\left(1-\frac{\phi_o^2}{2\phi^2_{\textrm{max}}}\right)\phi_o^2\,
\ee
It should be noted that this is a slight overestimation of the value of $\N$ but this does not constitute a problem for our analysis.
From this last equation, we can find the value $\phi_{o}^\mathcal N$
of the scalar field at the onset of inflation, for a given value of $\mathcal N$ as
\be
\phi_{o\pm}^\mathcal N=\pm\frac{\sqrt{3(1-\sqrt{1-8\mathcal N\gamma^2\lambda^2m^2/3})}}{\sqrt{8\pi}\gamma\lambda m}
\label{phio}
\ee
In the GR limit, that is, in the  $\lambda\mapsto 0$ limit, we expect that $\phi_{o\pm}^\mathcal N$ be equal to $\pm\sqrt{\mathcal N/2\pi}$.

Let us now see how we can find $\phi_B^o$ which is the value of the scalar field at the bounce that evolves under the dynamics to $\phi_{o\pm}$ as the starting point of inflation. According to Eq.(\ref{phi}), if at the bounce $\phi_B^o>0$ then $\ddot\phi_B<0$ (and $\dot\phi>0$). Similarly, if $\phi_B^o<0$ then $\ddot\phi_B>0$ (and $\dot\phi_B>0$). In the second case, after some time, $\ddot\phi$ becomes zero and after that it will be negative, but near the onset of inflation it becomes zero again. Near the start of inflation at the time for which $\dot\phi=0$, the value of the scalar field is larger than $\phi_B^o$. After that, $\dot\phi$ becomes negative and the value of the scalar field starts to decrease but very soon after $\dot\phi=0$, the inflationary era starts and the scalar field at the onset of inflation remains larger than the value of the scalar field at the bounce ($\phi_B^o<\phi_{o\pm}$).

Furthermore, because of the uniqueness of the solutions, $\phi_o$ is a monotonic function of $\phi_B$ and since $\phi_o$ is always greater than $\phi_B$, then it is an increasing function of $\phi_B$.

Given this, we can write the probability of having inflation with $\mathcal{N}$ e-folding or more as the quotient
of the volume on the space of solutions occupied by solutions with $\N$ or more e-foldings divided by the total
volume. Since the measure does not depend on volume $v$ and the range of this coordinate is infinite, both terms
are unbounded. However, we can very easily get rid of these spurious infinities by an appropriate renormalization 
(or gauge fixing \cite{thesis}). One possibility is to restrict the domain of the volume integral to the 
interval $v\in(1,2)$ (in Planck units). With this choice, the volume integrals in the quotient cancel 
each other and we get,
\ba
P_\mathcal{N} &=& \frac{\int^{\phi_{a}^{\mathcal{N}}}_{-\phi_{\textrm{max}}}\sqrt{1-F_B}\,\textrm{d}\phi_B+\int_{\phi_{b}^{\mathcal{N}}}^{\phi_{\textrm max}}\sqrt{1-F_B}\,\textrm{d}\phi_B}{\int_{-\phi_{\textrm{max}}}^{\phi_{\textrm{max}}}\sqrt{1-F_B}\,\textrm{d}\phi_B}
\nonumber\\
& = & 1-\frac{\int^{\phi_{b}^{\mathcal{N}}}_{\phi_{a}^{\mathcal{N}}}\sqrt{1-F_B}\,\textrm{d}\phi_B}{\int_{-\phi_{\textrm{max}}}^{\phi_{\textrm{max}}}\sqrt{1-F_B}\,\textrm{d}\phi_B}\, ,
\ea
where $\phi_{a}^{\mathcal{N}}$ and $\phi_{b}^{\mathcal{N}}$ are the minimum and maximum value of $\phi$ at the bounce that cause inflation with $\mathcal N$ e-folding respectively and $\phi_{\textrm{max}}$ is the maximum value of $\phi_B$ and is equal to  $3/2\gamma\lambda m\sqrt\pi$. Then
\be
\begin{split}
P_{\mathcal{N}}&=1-\frac{\arcsin(2\gamma\lambda m\phi_{b}^{\mathcal{N}}\sqrt{\pi}/3)-\arcsin(2\gamma\lambda m\phi_{a}^{\mathcal{N}}\sqrt{\pi}/3)-2\gamma\lambda m\sqrt{\pi}/3(\phi_{b}^{\mathcal{N}}-\phi_{a}^{\mathcal{N}})}{2(\pi/2-1)}\\
\end{split}
\label{pp}
\ee

We have plotted in Fig.~\ref{Fig:1} the dynamical trajectories for three values of $\lambda$. As one can see, when $\lambda$ becomes small, the trajectories (for finite values) are almost parallel. Then, in the limit $\lambda\to 0$ we can approximate $\phi_b^\mathcal N-\phi_a^\mathcal N$ by $\phi_{o+}^\mathcal N-\phi_{o-}^\mathcal N$ and since $\phi_{o\pm}^\mathcal N$ are finite, then the difference between $\phi_b^\mathcal N$ and $\phi_a^\mathcal N$ is finite. From the above discussion and Eq.(\ref{pp}), for a finite $\mathcal N$ we see that the probability is a decreasing function of $\lambda$ ($2\gamma\lambda m\sqrt{\pi}\phi_b/3$ and $2\gamma\lambda m\sqrt{\pi}\phi_a/3<1$) and when $\lambda$ goes to zero we have that $\arcsin(2\gamma\lambda m\phi_{b}^{\mathcal{N}}\sqrt{\pi}/3)\approx 2\gamma\lambda m\phi_{b}^{\mathcal{N}}\sqrt{\pi}/3$ (and equivalently for $\phi_a^\mathcal N$) and therefore the probability in Eq.(\ref{pp}) goes to 1.  This is the first result of this paper.

Let us now understand qualitatively why the probability increases as the LQC parameter decreases. As the analysis here presented and that of \cite{ash:sloan} shows, for a given value of $\rho_{\textrm{crit}}$, there is an interval
$(\phi_a^{\cal N},\phi_b^{\cal N})$, in the `kinematically dominated regime' (where the energy density at the bounce is mainly due to the kinetic energy), where there are not enough e-foldings. This interval, as we have estimated before, depends on $\lambda$. In Fig.~\ref{Fig:3} we have plotted, for three values of $\rho_{\textrm{crit}}$, the `critical trajectories' for which the transition occurs. That is, these trajectories have an almost identical behavior at small densities, so they inflate in the same fashion, and touch the bounce surface at the
points  $\phi_a^{\cal N}$ and $\phi_b^{\cal N}$. If we now follow them to higher densities `back in time', what one sees from the graph is that as $\lambda$ decreases, and
$\rho_{\textrm{crit}}$ increases, the intersection points tend to the $\dot{\phi}$ axis. The relative size of the interval $(\phi_a^{\cal N},\phi_b^{\cal N})$ in the total allowed interval $(-\phi_{\textrm{max}},\phi_{\textrm{max}})$ also goes to zero as $\rho_{\textrm{crit}}\to \infty$\footnote{Recall that $\phi_{\textrm{max}}$ is obtained from the value of the potential at the bounce $\rho_{\textrm{crit}}=V(\phi_{\textrm{max}})$. In our case $\phi_{\textrm{max}}=\sqrt{2\,\rho_{\textrm{crit}}}/m$, so the interval in which the scalar field can take values also diverges as $\rho_{\textrm{crit}}\to \infty$.}. Since the integrand does not diverge, this already implies that the quotient vanishes and the probability goes to 1. 

Note also that this result is independent of the precise value of $\N$ (as long as it is large enough for our approximation to be valid). Does this mean that we can take the limit $\N\to\infty$ and also have probability one? In order to answer this one should exercise some care. For any finite value of $\N$, the probability will tend to one as we make $\l$ smaller for two reasons. The first one is that the dynamics of the effective equations is such that those trajectories that do not have enough inflation get `funneled', for large enough values of the critical density, into the interval $(\phi_a^{\cal N},\phi_b^{\cal N})$ that remains bounded, while the total interval for $\phi$ grows with $\rho_{\textrm{crit}}$. This only happens because we are taking the bounce surface as the reference surface where the probability is computed. Furthermore, the measure is such that relative volume we associate to those trajectories is very small and becomes zero in the $\l\to 0$ limit. This does not mean that we can fix $\l$ and, say, take the limit $\N\to\infty$.

A final remark is in order. In our analysis, as plotted in Fig.~\ref{Fig:3}, the criteria for how much inflation there is coincides with that of \cite{ash:sloan}.
That is, we start with the critical density of LQC (of the order of the Planck density), which gives an initial condition from which to measure e-foldings, and find those trajectories --in theories with a different $\l$-- for which the dynamics at low densities coincide, where inflation actually occurs. This is also in the spirit of \cite{BKGZ}, which suggested to take initial conditions at the Planck scale.  
Our strategy has to be contrasted with a possible alternative that involves going closer to the big bang, as we decrease $\l$, and use that as initial condition in the e-folding counting. The problem with this choice is that, as one approaches the big bang that has zero volume, the number of e-foldings diverges for all trajectories, so even the question of which trajectories have enough inflation becomes meaningless, since every trajectory would have an infinite number of e-foldings.  

\begin{figure}[htb]
\centerline{\includegraphics [scale=0.4]{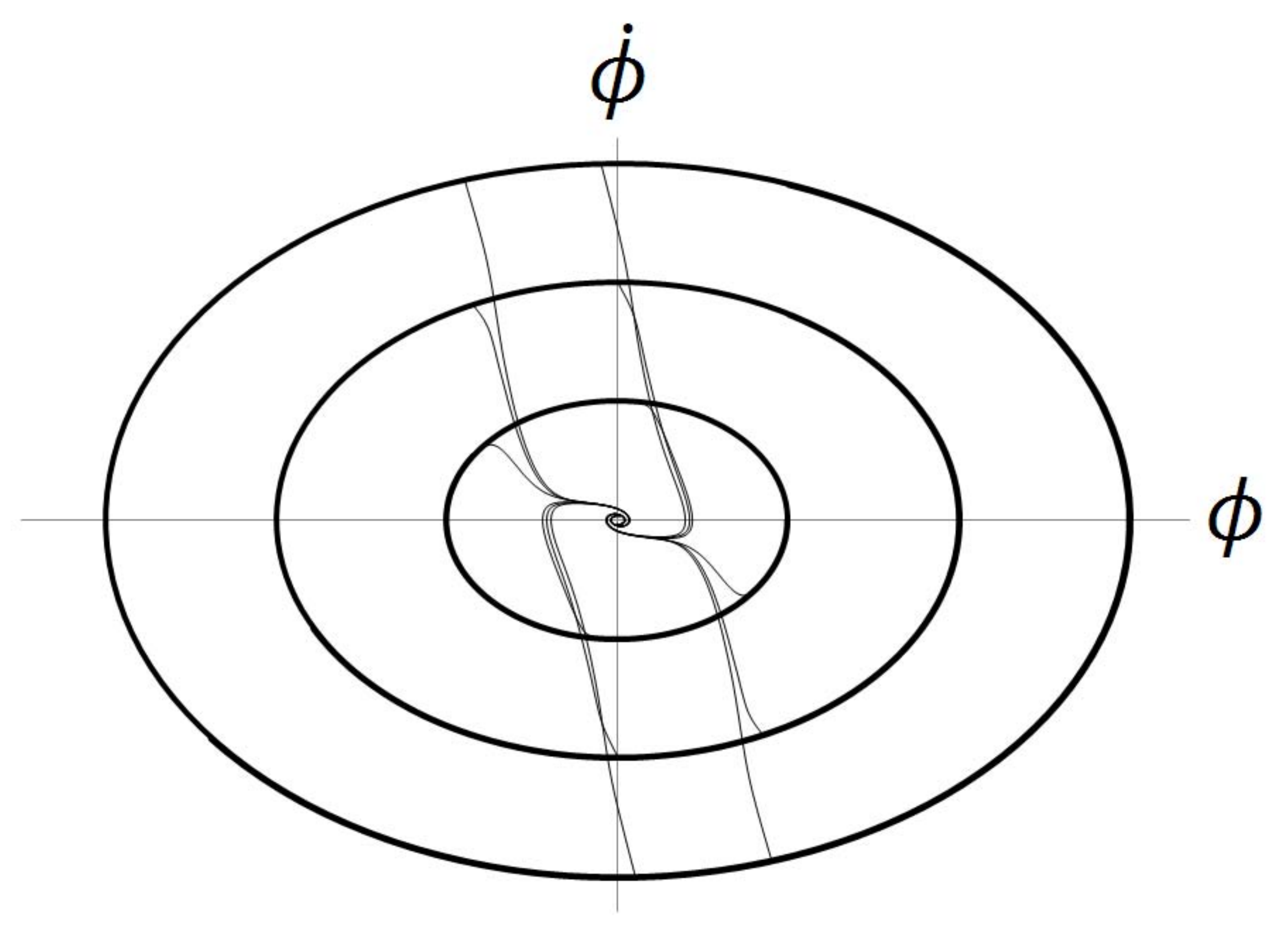}}
\caption{Here we are plotting trajectories for three different values of the critical density $\rho_{\textrm{crit}}$. In each case, we have a boundary of the trajectories in the $(\dot\phi,\phi)$ plane, corresponding to the bounce, that are depicted as ellipsoids. The smallest ellipsoid can be taken as the LQC one, and the larger ones are closer to the GR limit. The `critical' curves that separate the region of enough e-foldings, as determined by the LQC scale, are then plotted for the three different values of the critical density $\rho_{\textrm{crit}}$. One can see that, as the critical density increases, the intersection with the ellipsoid of critical density comes closer to the $\dot{\phi}$ axis. }
\label{Fig:3}
\end{figure}

\subsection{Comparison}

Let us now come to the question of how we can reconcile the results of Gibbons and Turok  \cite{gibb:turok} on the one side and those of Ashtekar and Sloan \cite{ash:sloan} on the other. The first possible objection is: How can we compare two results that are taken on two different theories, GR on one side and LQC on the other? As we have seen before, one can in fact approximate very well the low density GR trajectories by (low density) LQC effective trajectories.  Thus, the region of interest in the Gibbons and Turok analysis, $\rho_{\textrm{GT}}/\rho_{\textrm{crit}}\ll 1$, which is for trajectories near the end of inflation (and therefore, around the constant density surface in our Figure~\ref{Fig:3}), one can take the LQC effective trajectories without any problems as a very good approximation to the GR dynamics. This allows us to `embed' the low density GR dynamics in the effective LQC description with very good accuracy.

With this assumption, we can now compare the two result within the effective LQC description. We have two constant density surfaces, as depicted in Fig.~\ref{Fig:4}. The external ellipsoid corresponds of course to the critical density $\rho_{\textrm{crit}}$ at scale $\lambda$, while the small one corresponds to the density $\rho_{\textrm{GT}}$ as chosen by Gibbons and Turok\footnote{The figure is not to scale, since we are asking that $\rho_{\textrm{GT}}\ll \rho_{\textrm{crit}}$. The relative densities in the figure were taken to illustrate our point.}. One puzzling fact about the huge discrepancy in results is that both analysis use the natural Liouville measure (properly normalized) to compute the probability on constant density surfaces. One important property of the Liouville measure is that it is invariant under the dynamical evolution. So, how come we arrive to two very different conclusions?

There are two key observations to understand this apparent tension. 
The first one pertains to the question of whether the time evolution invariance of the Liouville measure implies that the probability is also invariant. On a first view, one might imagine that the probability has to be invariant since one is just measuring the relative phase space volume of those
trajectories with $\N$ e-foldings or more, relative to the total volume in phase space. Now, the technical step that allowed to normalize the phase space volume (the total phase space volume is infinite) in \cite{gibb:turok} and \cite{ash:sloan} was to realize that there is an invariance in the space of classical solutions by rescaling the physical volume. This invariance has its origin in the fact that, instead of describing the whole universe, one has to restrict attention to a fiducial region $R$ in space (the spatial volume of the whole universe in $k$=0 FRW is infinite, so  one needs to consider a region with a finite volume). Since this choice is arbitrary, one can in principle chose a smaller/larger region for which we assign a smaller/larger volume, but the physics should be unchanged. When one takes care of this ambiguity, either by taking an appropriately chosen `interval in $v$' as we done in the previous part, or by an appropriate gauge fixing \cite{ash:sloan,thesis}, one still has to be careful about the possible change in physical volume during the dynamical evolution that would also induce a change in relative volume in phase space.

Let us see how this comes about. 
Invariance of the Liouville measure means that the volume in phase space is preserved. Let focus our attention in the quadrant in the space of solutions, with coordinates $(v_B,\phi_B)$, defined by $1 \leq v_B\leq 2$ and $-\phi_{\textrm{max}}\leq \phi_B\leq \phi_{\textrm{max}}$, and follow it through its dynamical evolution. If we now
take another `gauge fixing' at a lower energy density, say $\rho_1$ (See Fig.~\ref{Fig:4}), we immediately notice that
the range in $\phi$ is much smaller. Since the total volume of the quadrant we are following has to be the same (due to the dynamical invariance of the Liouville measure), the range in $v$ has to increase, as it indeed does, since most solutions inflate. The crucial point here is to realize that the change in volume $\Delta v=v_1-v_B$ from the bounce to the $\rho_1$ surface depends on the value of $\phi$. Thus, the lines, say, $v_B(\phi_B)=1$ at the bounce gets mapped,
in general to a curve $v_1(\phi_1)$ that is no longer constant as a function of $\phi_1$. That is, each solution has a different change in physical volume depending on the value of the scalar field at the bounce. But,
if one is only keeping track of the change in the  `phase space  coordinate' $\phi$ when computing the probability, then the {\it relative volume in phase space}, as measured by only $\phi$, can indeed change. Since this is precisely what one means by probability in the analysis of \cite{gibb:turok,ash:sloan} and here, we are led to conclude that the probability  indeed depends on the surface on which it is computed. Since this argument did not use any particular detail of the LQC dynamics, this ambiguity in the probability depending on the choice of constant density surface is also present in general relativity. Let us now see what further assumption are made in both calculations.

 \begin{figure}[htb]
\centerline{\includegraphics [scale=0.4]{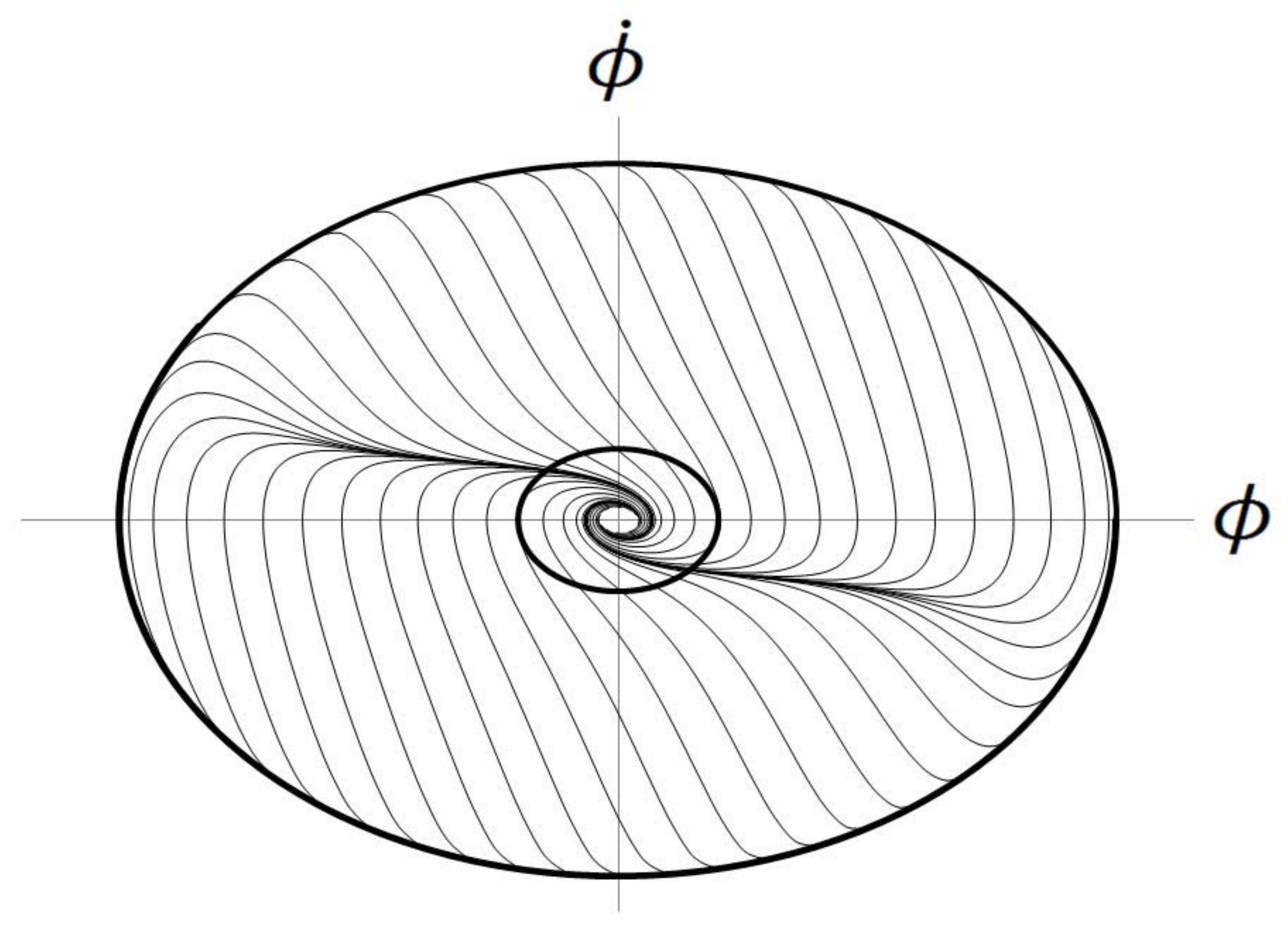}}
\caption{For a fixed value of $\rho_{\textrm{crit}}$, we plot the exterior, critical density surface and a surface of constant density $\rho_{\textrm{GT}}\ll \rho_{\textrm{crit}}$ (not drawn to scale, of course) on the $(\dot\phi,\phi)$ plane. Trajectories with a uniform distribution at the LQC bounce ellipsoid  are plotted. 
Note that trajectories for which there is enough inflation get funneled into a small region in the
smaller $\rho_{\textrm{GT}}$ ellipse. Near this surface, the GR and LQC dynamics almost coincide}
\label{Fig:4}
\end{figure} 

The second observation is the following.
When computing the probability of having
$\N$ e-foldings, one has to assume an a-priory probability distribution ${\mathbf{P}}(\phi,v)$ of the classical trajectories, and then integrate this probability distribution with respect to the corresponding measure. In \cite{ash:sloan}, the authors consider the most natural choices, namely, the probability is computed on the critical density surface (i.e., the bounce) using, as the integration measure, the Liouville measure. By invoking Laplace's `principle of indifference' as in \cite{gibb:turok}, they consider
 a {\it uniform} distribution on the
space of trajectories (labeled by $(\phi,v)$) and performed an appropriate gauge fixing with respect to the
volume rescaling freedom available, in the same spirit we have done here.
 We have illustrated this scenario in Fig.~\ref{Fig:4}, where we plotted trajectories uniformly distributed in $\phi$ along the critical density surface. 
If we now follow these trajectories along the dynamical evolution we notice that, when they intersect the 
$\rho=\rho_{\textrm{GT}}$ surface, they are no longer uniformly distributed. Quite the opposite. Due to the
global properties of the dynamics, the trajectories are funneled into the `attractor' on the plane $(\phi,\dot{\phi})$ and, therefore, effectively acquire a 
new `probability distribution' $\tilde{\mathbf{P}}(\phi)$ on the  $\rho=\rho_{\textrm{GT}}$ surface\footnote{We can
view this induced probability distribution as a way of keeping track of the relative change in phase space volume do to the dynamics that induces a differentiated change in physical volume $v$ for different trajectories. Namely,
the distribution is given by $\tilde{\mathbf{P}}(\phi)=v_{\textrm{GT}}(\phi)$, in terms of the volume $v_{\textrm{GT}}$, as a function of $\phi$ on the Gibbons-Turok constant density surface.}. 

If we were to compute the probability of inflation on the Gibbons and Turok surface, but {\it weighted with the induced distribution} $\tilde{\mathbf{P}}(\phi)$, we would of course get the same result of Ashtekar and Sloan, given our previous discussion. What Gibbons and Turok did instead was to assume a uniform distribution ${\mathbf{P}}(\phi)$ on the $\rho=\rho_{\textrm{GT}}$ surface. With respect to the uniform distribution, the phase space volume of inflating solutions is very small and the probability is therefore, very close to zero. Had we chosen to compute the probability on a surface with even lower density, the result would even be smaller. This constitutes the main difference between both calculations\footnote{Note that Linde had pointed out that the assumption of Gibbons and Turok to take uniform initial conditions at the end of inflation might be the source of their negative result  \cite{linde}, but the mechanism he outlined is different from ours, since he was not taking the Liouville measure into account.}.

Furthermore, we can now understand why the probability found by Gibbons and Turok is so small. If we look at the region of Fig.~\ref{Fig:3} for which there is not enough inflation (in between the critical curves) on the LQC critical density surface, and follow those trajectories as in Fig.~\ref{Fig:4}, we see that those trajectories occupy now a much larger region on the $\rho=\rho_{\textrm{GT}}$ ellipsoid. In other words, the trajectories for which there
is enough inflation get funneled into a small region in the $\rho=\rho_{\textrm{GT}}$ surface that, when integrated with respect to the uniform distribution ${\mathbf{P}}(\phi)$  of  \cite{gibb:turok}, give a very small contribution to the probability.  
One could also consider the opposite situation in which one starts with an uniform distribution on the GT surface and `evolve back' in time to the bounce surface. In that case, the dynamics
will `expel' the trajectories in such a way that the probability distribution ${\mathbf{P}}'(\phi)$  
induced on the bounce surface
is concentrated on the region where there is not enough inflation. If one integrates that probability distribution with respect to the Liouville measure the resulting probability is very close to zero, as found by Gibbons and Turok in \cite{gibb:turok}.\footnote{One should note that, as previously discussed, in order to make the distinction of which trajectories have enough e-foldings, one has to introduce a cut-off for the initial condition. Here we have adopted the LQC scale as a natural unambiguous choice.} 

\section{Discussion}
\label{sec:4}

Let us summarize our results. We have reanalyzed the treatment of the simplest inflationary model from the perspective of loop quantum cosmology. By using effective equation we studied the structure of the space of classical solutions with the aim of answering the question: How probable is it to achieve enough e-foldings? 
In particular we have considered this question keeping the discreetness parameter of loop quantum cosmology as a free parameter. When the parameter vanishes, one expects the dynamics to reduce to the standard, general relativity behavior. 
The first result is that, as previously shown in \cite{ash:sloan}, the probability for enough inflation is very close to one when the discreetness parameter $\l$ is of the order of the Planck scale. We then considered the dependence of the probability as one decreases the parameter and it approaches the general relativity limit. As we have shown, the probability increases and approaches one as one reaches the limit. Next, we studied the global properties of the system to understand the underlying reason for the discrepancy of these results and those of \cite{gibb:turok} in which the probability of enough inflation was computed to be close to zero, within general relativity. What we found is that this discrepancy is due to the differences in the underlying assumptions in both calculations. As it turns out the probability as computed in both \cite{gibb:turok} and \cite{ash:sloan} depends very strongly on the
{\it constant density surface} where it is calculated.  While Ashtekar and Sloan assume a uniform distribution of classical trajectories at the naturally defined surface available due to the universal existence of the bounce, Gibbons and Turok take it at an arbitrarily defined surface at the end of inflation. Given the large difference in scales involved and due to the global properties of the dynamics and the probability measure,  these two assumptions have strikingly different consequences. During the evolution from the bounce to the Gibbons-Turok scale, most of the trajectories that undergo enough inflation --contributing significantly to the probability-- get funneled into a small region at the later scale, that has a correspondingly small contribution to the probability. This is the origin of the apparent tension. 

We have thus found two very different results even for GR. On the one hand the Gibbons-Turok result involves several, somewhat ad-hoc, assumptions given that there is no preferred choice of scale on GR. On the other hand, there is the limit of LQC when the discreetness parameter vanishes. In this later case we have, for each scale, calculations based on unambiguous and natural choices that provide a well defined result, even when the GR limit of LQC is non-smooth.
Thus even when in the $\l\to 0$ limit one is approaching arbitrarily close to the singularity, the probability of having $\N$ e-foldings --as measured from the Planck scale down-- can be given some meaning. As we have seen, the result that in the GR, $\l\to 0$ limit, the probability goes to one for any finite value of $\N$, seems to be generic. Whether this result is physically meaningful is, however, a completely different issue. In the situation in which $\lambda$ is taken well below the Planck scale, we are implicitly assuming that the classical equations are still valid. This is perhaps, too strong an assumption. One generically  expects quantum effects to dominate near and below the Planck scale. This is precisely what LQC provides for us via its effective equations.

Let us end with a series of remarks.  
\begin{enumerate}
\item Given that these two results are based on assumptions that yield completely opposite predictions, one might then ask what is the physically reasonable assumption to make? How can we justify one choice over the other? Is there a `canonical' choice of initial condition? In loop quantum cosmology we know that the bounce is generic for inflationary potentials \cite{massive}, and the effective equations are a very good approximation to the dynamics of semiclassical states \cite{victor}. Since every such effective trajectory goes through a bounce, selecting the surface of constant density, at the bounce, seems a rather natural choice. One should emphasize then that there does not exist a similar surface that is preferred in the classical GR case. As we have seen, the probability does depend in a rather dramatic way on the choice of such surface. Without any extra input, the LQC choice seems to be the most natural.

\item Even if one does not regard loop quantum cosmology as a fundamental theory, one can still view its effective dynamics and choice of surface as in \cite{ash:sloan} as a procedure to {\it regulate} the classical calculation. The bounce provides then the preferred `cut-off' surface envisioned by the authors of \cite{BKGZ}, but in an unambiguous fashion. From this perspective, what is amazing is that one can {\it remove the regulator} and obtain a finite answer. Furthermore, this `canonical answer' indicates that inflation with enough e-foldings is generic even in this particular way of approaching the GR limit.

\item As we have seen from our analysis here, the reason for the LQC result stems from the choice of surface where to compute the probability and {\it not} directly from any effect from the quantum geometry underlying LQC. In fact, as we remove the parameter encoding the quantum geometric effects, the probability of inflation increases\footnote{This is consistent with early calculations on LQC where the integration was performed at low densities \cite{germani}.}.

\item In our analysis we have used qualitative aspects of the global dynamics of the system. Therefore, our arguments and conclusions are insensitive to changes in the free parameters of the model. That is, our result are rather robust.

\item One should keep in mind that these results are purely classical. It is to be expected that quantum effects might provide a more realistic {\it distribution} on the space of classical trajectories. Some proposals have been put forward, but in the context of particular states and only for those trajectories satisfying WKB conditions \cite{har:her}. One could imagine that semiclassical states in LQC might provide an improved distribution from which one might get a `quantum corrected' estimation of the probability for enough inflation. This matter should certainly be studied in detail.

\end{enumerate}

\section*{Acknowledgments}

\noindent We would like to thank I. Agullo, A. Ashtekar, D. Sloan and P. Singh for 
discussions and comments. This work was in part supported by DGAPA-UNAM IN103610 grant, by NSF
PHY0854743 and by the Eberly Research Funds of Penn State.

\end{document}